\documentclass[rnote]{aa} 
\usepackage{epsfig}
\usepackage{rotating}
\usepackage{natbib}
\usepackage{graphicx}
\usepackage{graphics}

\newcommand{\hst}{{\sl HST}}

\newcommand{\vlt}{{\sl VLT}}
\newcommand{\fors}{{\sl FORS1}}
\newcommand{\vltn}{{\sl Very Large Telescope}}

\newcommand{\chan}{{\sl Chandra}}
\newcommand{\vlbi}{{\sl VLBI}}

\newcommand{\fo}{\ensuremath{f^\mathrm{o}}}
\newcommand{\fe}{\ensuremath{f^\mathrm{e}}}
\newcommand{\pu}{\ensuremath{P_{\rm U}}}
\newcommand{\pq}{\ensuremath{P_{\rm Q}}}
\newcommand{\pl}{\ensuremath{P_{\rm L}}}
\newcommand{\pli}{\ensuremath{P_{\rm L}^{\rm intr}}}
\newcommand{\plm}{\ensuremath{P_{\rm L}^{\rm mod}}}

\def\gasm{\hbox{\hspace{0.5mm}}\raise2pt
       \vbox{\hbox{$>$}}\lower2pt
              \vbox{\moveleft6.0pt\hbox{$\sim$ }}\hbox{\hskip 0.05mm}}

\begin{document}

  \title{The Optical Polarisation of the Vela Pulsar revisited\thanks{Based on observations collected at the European Southern Observatory, Paranal, Chile under  programme ID 63.P-0002(A) } }

 \author{R.P. Mignani\inst{1}
 \and 
	S. Bagnulo\inst{2}
\and 
	J. Dyks\inst{3}
\and 
	G. Lo Curto\inst{2} 
\and 
	A. S\l{}owikowska\inst{3,4}
}

   \offprints{R.P. Mignani}

   \institute{Mullard Space Science Laboratory, Holmbury St. Mary, Dorking - Surrey, RH5 6NT United Kingdom\\
              \email{rm2@mssl.ucl.ac.uk}
\and
European Southern Observatory, Alonso de Cordova 3107, Vitacura, Santiago, Casilla 19001 Santiago 19, Chile\\ \email{sbagnulo@eso.org, glocurto@eso.org}
\and
Nicolaus Copernicus Astronomical Center,  Rabia\'nska 8, 87-100 Toru\'n, Poland \\
\email{jinx@ncac.torun.pl},\email{aga@ncac.torun.pl}
\and
 Max-Planck-Institut f\"ur Extraterrestrische Physik, Giessenbachstrasse, Garching, D-85748 Germany   \\ \email{aga@mpe.mpg.de} }

     \date{Received ...; accepted ...}

   \abstract{}{In this  work we present  a revised measurement  of the
phase-averaged optical polarisation of the Vela pulsar (PSR B0833-45),
for which only one value has  been published so far (Wagner \& Seifert
2000).  }{Our measurement has been obtained through an accurate reanalysis
of  archival  polarisation   observations  obtained  with  the  \fors\
instrument at  the \vlt.  }{We  have measured a  phase-averaged linear
polarisation degree $\pl=9.4\% \pm 4\%$ and a position angle $\theta =
146^\circ  \pm  11^\circ$, very  close  to the  ones  of  the axis  of
symmetry of  the X-ray  arcs and  jets detected by  \chan\ and  of the
pulsar  proper motion.}{We have  compared the  measured phase-averaged
optical  polarisation  with  the  expectations of  different  pulsars'
magnetosphere  models.  We have  found  that  all models  consistently
predict  too large  values of  the phase-averaged  linear polarization
with respect to  the observed one.  This is  probably a consequence of
present models' limitations which neglect the contributions of various
depolarisation  effects.  Interestingly,  for the  outer gap  model we
have  found  that,  assuming  synchrotron radiation  for  the  optical
emission,  the observed  polarisation position  angle also  implies an
alignment between the pulsar rotational axis and the axis of symmetry  of the X-ray arcs and jets.  }

   \maketitle

\section{Introduction} 
Polarisation measurements of pulsars  and of their synchrotron nebulae are
uniquely  able to  provide deep  insights into  the  highly magnetised
relativistic environment of young rotating neutron stars.  Besides the
radio band, optical observations  are primarily suited to provide such
insights.  Being the first, and the brightest ($V \sim 16.5$), optical
pulsar  identified so  far, it  comes as  no surprise  that  the first
optical polarisation  measurements were obtained for the  Crab pulsar (PSR
B0531+21).  Strong polarisation is  actually expected when the optical
emission is  produced by synchrotron  radiation which, in the  case of
the Crab  pulsar, is certified  by its power-law spectrum  (see, e.g.,
Sollerman et al.~ 2000).  Indeed,  the polarisation of the Crab pulsar
optical emission was discovered (Wampler  et al.~ 1969) soon after its
optical  identification  (Cocke   et  al.~  1969).   Accurate  optical
phase-resolved polarisation measurements of the Crab pulsar (e.g. Smith et
al.~ 1988;  Kanbach et al.~ 2005)  have shown the  phase dependence of
the polarisation degree, with a maximum  of $\sim 50 \%$ in the bridge
between   the   main  and   the   interpulse.   Optical   polarisation
measurements   are  extremely   useful   in  providing   observational
constraints   to  test   the   different  models   proposed  for   the
magnetospheric activity of pulsars.  Unfortunately, despite the number
of pulsars detected in the optical band has increased significantly in
the last  ten years  (see Mignani  et al.  2004  and Mignani  2005 for
updated reviews), the Crab pulsar is still the only one for which both
precise and  repeated polarisation  measurements have been  obtained.  For
the other young pulsars (age  $\le$ 10\,000 years) only preliminary or
uncertain phase-averaged  optical polarisation measurements  were reported
so far.   For the second  youngest ($\sim 1600$ years)  optical pulsar
PSR  B0540-69  ($V\sim 22.5$),  considered  the  Crab pulsar  ``twin''
because  of its  overall similarities  in age,  period  and energetics
(e.g. Caraveo et al.~ 2001a), polarisation observations with the \vlt\
were reported by Wagner  \& Seifert (2000).  However, the polarisation
measurement   ($\approx$  5\%   with  no   quoted  error)   was  certainly
contaminated by the contribution  of the compact ($\sim$ 4'' diameter)
surrounding  synchrotron  nebula (Caraveo  et  al.~  2001a).  For  PSR
B1509-58,  the  next  youngest  ($\sim  5000$  years)  pulsar  with  a
candidate   optical   identification,  the   value   of  the   optical
polarisation  is  also   very  uncertain.   The  original  counterpart
($V=22$) proposed by Caraveo et  al. (1994) was discarded by Wagner \&
Seifert  (2000)  on   the  basis  of  the  lack   of  evident  optical
polarisation in favour  of a much fainter ($R  \sim 26$) object hidden
in the PSF  wings of the star, whose detection  has been now confirmed
by  IR  observations  (Kaplan   \&  Moon  2006).   However,  both  the
difficulties  in  the  PSF   subtraction  and  the  faintness  of  the
counterpart make  the reported polarisation measurement  ($\approx 10$ \%,
also quoted with no error) only tentative.  A preliminary polarisation
measurement was  reported by  Wagner \& Seifert  (2000) also for  the Vela
pulsar  (PSR  B0833-45),  the  third  brightest ($V  \sim  23.6$)  and
youngest  ($\sim 10\,000$  years)  optical pulsar.   Finally, for  the
older   ($\sim   100\,000$  years)   PSR   B0656+14   $(V  \sim   25$)
phase-resolved  optical polarisation  measurements have  been  obtained by
Kern et  al.  (2003).  Like for  the Crab, also  for PSR  B0656+14 the
polarisation degree is phase-dependent, with a maximum ($\sim$ 100 \%)
in the bridge between the two  pulses.  \\ For the Vela pulsar, Wagner
\& Seifert (2000) published only  the value of the linear polarisation
degree.   The  need to  obtain  a  more  complete description  of  the
polarisation properties  of Vela pulsar by including  e.g., the linear
polarisation  position angle  has  prompted us  to  undergo a  careful
reanalysis  of   the  original  data   set.   The  knowledge   of  the
polarisation position angle is indeed an additional important piece of
information  to  test different  models  on  the  optical emission  of
pulsars.   This  paper is  organised  as  follows: observations,  data
analysis and  results are described  in Sect.~2, while  the comparison
with existing pulsar emission models is discussed in Sect.~3.

\section{Observations and Data Analysis}

\subsection{Observations}

 Phase-averaged  polarimetry  observations  of  the Vela  pulsar  were
obtained  at the first  Unit Telescope  (UT1) of  the ESO  \vltn\ with
\fors\   on   April   12   1999   as  a   part   of   the   instrument
commissioning\footnote{see
http://archive.eso.org/archive/eso\_data\_products.html}.        \fors\
(FOcal Reducer Spectrograph), a  multi-mode instrument for imaging and
(long-slit/multi-object)   spectroscopy  equipped   with  polarimetric
optics, is described in Appenzeller et al.  (1998).  Observations were
performed  in Standard Resolution  (SR) mode  (0\farcs2/pixel) through
the  $R$   Bessel  filter  ($\lambda=6570$\,\AA;   $\Delta  \lambda  =
1500$\,\AA),  with an average  seeing of  $\sim 0\farcs7$  and airmass
varying  between 1.4  and 1.9.   The instrument  set-up was  chosen to
allow for linear polarisation measurements.   A total of 6 frames were
taken with  the instrument position  angle (hence the  Wollaston axis)
oriented at  $+10^\circ$ North to  East. One 1\,000  s plus two  380 s
exposures were obtained  with the fast axis of  the half-wave retarder
plate parallel  to the Wollaston prims (rotation  angle $0^\circ$) and
the remaining  three (1\,000 s each) at  rotation angles $22.5^\circ$,
$45^\circ$, and $67.5^\circ$,  respectively (see Table 1 for  a log of
the observations).  In this work we  have used only the longest of the
exposures performed with the  $0^\circ$ retarder plate angle.  Science
data and  the associated calibrations have been  retrieved through the
ESO  archive  interface\footnote{http://archive.eso.org/} and  reduced
following the standard approach for de-biasing and flat-fielding.  The
master flat field has been obtained from the median-combination of sky
images obtained  during twilight with  no polarimetric optics  in.  No
observations of polarimetric standard  stars were performed during the
night, but analysis  of the polarimetric data taken  in the context of
the polarimetric calibration plan  shows that the instrument is stable
(Fossati et  al 2007).  We  have finally obtained four  reduced images
corresponding to  the four retarder  plate angles.  As an  example, we
show in  Fig. 1  the reduced image  taken with the  $0^\circ$ retarder
plate  angle.    We  have  checked   for  the  presence   of  extended
polarisation  features which  could be  spatially associated  with the
X-ray nebula detected  around the pulsar by \chan\  (Pavlov. 2001) but
we  have  found  none,  thus  confirming the  quick  analysis  briefly
reported in Mignani et al. (2003).

\begin{figure}
\centering           
\includegraphics[bb=28 35 328 335,width=8.0cm,clip]{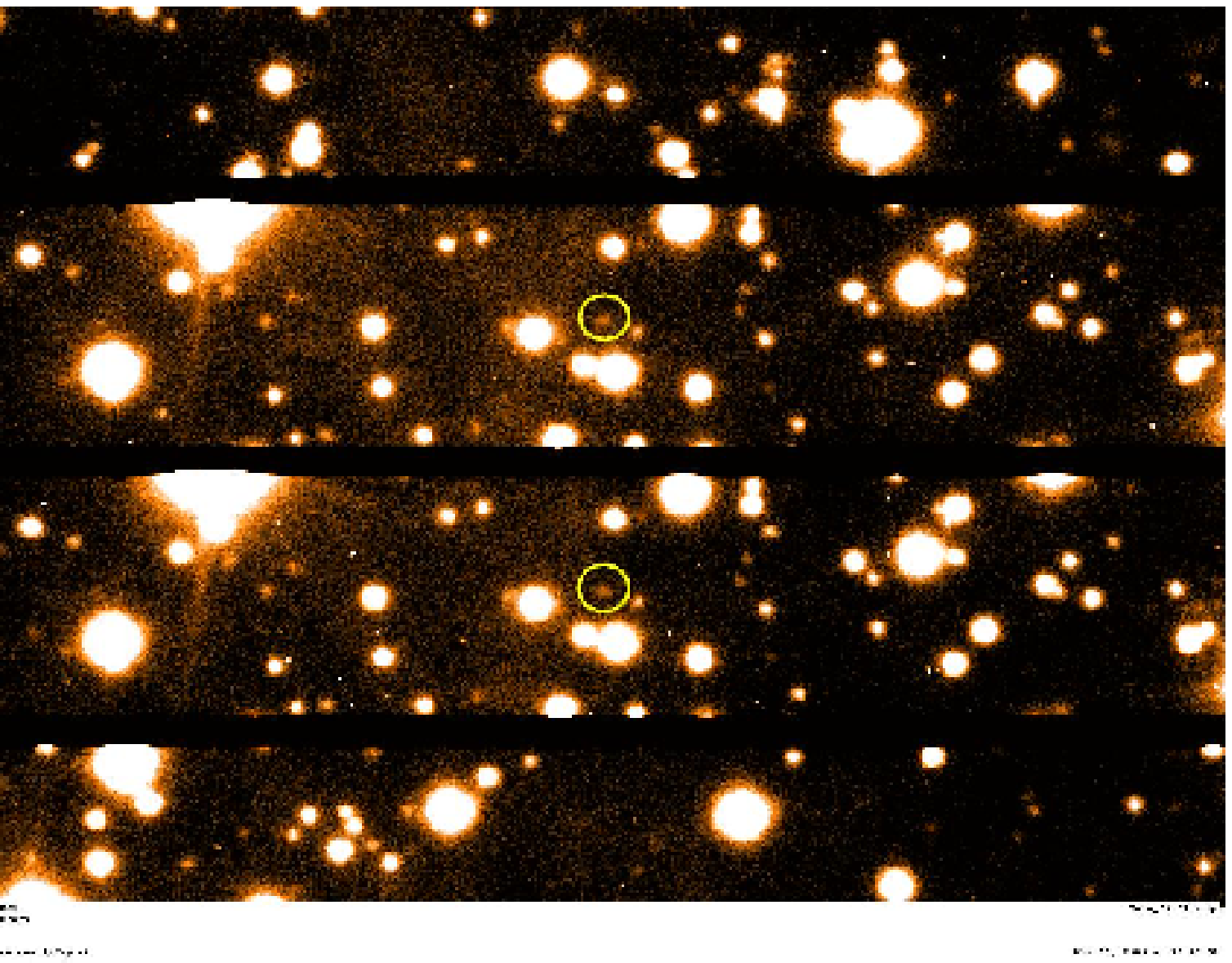}
\caption{VLT \fors\  image of  the Vela pulsar  field taken  with  the  $R$ 
Bessel  filter  and  polarised  optics, which split the source image in
ordinary and extra-ordinary beams. 
The pulsar counterpart ($R=23.9$) is highlighted by the circle.  }
\label{vela_fors}       
\end{figure}

\begin{table}
\begin{center}
\caption{Summary of  the polarimetric observations of  the Vela pulsar
field taken with FORS1.}
\begin{tabular}{ccccc} \hline
  Seq. & Angle($^\circ$) & Exposure (s) & Seeing (``) & Airmass	\\ \hline
1 & 0      & 1000 & 0.81 & 1.37 \\
2 & 22.5   & 1000 & 0.78 & 1.44\\
3 & 45.    & 1000 & 0.57 & 1.53\\
4 & 67.5   & 1000 & 0.62 & 1.63\\
5 & 0      & 380  & 0.61 & 1.76\\
6 & 0      & 380  & 0.62 & 1.90\\ \hline
\end{tabular}
\label{tabdatasummary}
\end{center}
\end{table}

\subsection{Data Analysis}

We  have  then  used  our   reduced  images  to  compute  the  linear,
phase-averaged polarisation  degree of the  Vela pulsar.  For  a given
source, the linear polarization degree \pl\ is calculated as:

$$\pl = (\pq^2 + \pu^2)^{1/2} \eqno{(1)} $$

\noindent
where, following the notations defined in Landi Degl'Innocenti et al. (2007):

 \[ \pq = \frac{Q}{I}\ \ {\rm and}\ \ \pu = \frac{U}{I} \]

\noindent
and $I$,  $Q$, $U$ are the  Stokes parameters defined  as in Shurcliff
(1962).  Applying  the recipe described  in Bagnulo et  al.(2006), the
\pq\  and \pu\  parameters  are computed  according  to the  following
relations:

$$
\pq  = 
\frac{1}{2}  \Bigg\{
\left(\frac{\fo - \fe}{\fo + \fe}\right)_{\alpha= 0\degr} -
\left(\frac{\fo - \fe}{\fo + \fe}\right)_{\alpha=45\degr}
\Bigg\}   \eqno{(2)} $$  

$$
\pu  = 
\frac{1}{2}  \Bigg\{
\left(\frac{\fo - \fe}{\fo + \fe}\right)_{\alpha=22.5\degr} -
\left(\frac{\fo - \fe}{\fo + \fe}\right)_{\alpha=67.5\degr}
\Bigg\} \eqno{(3)} $$

\noindent
where \fo\ and \fe\ are the source counts measured in the ordinary and
in the extraordinary mode, respectively  for each of the four retarder
plate angles  (0$^\circ$, 22.5$^\circ$, 45$^\circ$  and 67.5$^\circ$).
In  our  case, the  source  counts  have  been measured  via  standard
aperture photometry  using the photometric  package of IRAF  and using
optimised apertures for  both the source and the  background.  For all
retarder plate angles we have obtained photometric accuracies of $\sim
8$  \%, or  better.   To  minimise the  contamination  from the  three
brighter stars close to the pulsar  position (see Fig. 1) we have also
measured  the  source counts  through  PSF  photometry.  However,  the
results obtained with the two approaches do not differ significantly.

By  substituting the values  of \fo\  and \fe\  measured for  the Vela
pulsar for all of the four  retarder plate angles in Eqs.~(2) and (3),
we have thus obtained $\pq = 0.02\,\% \pm 3.7$\,\% and $\pu = -9.4\,\%
\pm 4.0$\,\%, where  the errors on \pq\ and \pu\  are given by Eq.~(3)
of Fossati  et al.\ (2007). These  values have been  then corrected as
explained in  the FORS  User Manual  and in Fossati  et al.  (2007) to
account for the chromaticity  of the retarder waveplate.  These values
are  measured in  the  detector reference  frame,  which was  oriented
$+10^\circ$ North to East (see  Sect.~2.1).  In order to refer them to
the celestial reference frame  we have applied a counter-rotation with
respect  to the  instrument  position angle,  using  Eq.~(9) of  Landi
Degl'Innocenti et al.   (2007) with $\chi = -10\degr$.  We thus end up
with $\pq'  = 3.6\,\% \pm 3.7$\,\%  and $\pu' =  -8.7\,\% \pm 4.0$\,\%
where $\pq'$  and $\pu'$ represent the linear  polarization measured using
the North Celestial Meridian as reference direction.

\subsection{Results}
By substituting $\pq'$ and $\pu'$ in Eq.~(1), we have finally obtained
$\pl = 9.4 \pm 4$\,\%,  where the associated error has been calculated
using Eq.~(6) of  Fossati et al. (2007).  This  relatively large error
is justified  by the fact that,  owing to the faintness  of the target
($R=23.9$),  the uncertainty on  the background  subtraction dominates
the photometric errors on \fo\  and \fe\ and ultimately those on $\pq'$
and  $\pu'$.   We have  checked  that our  measurement  is  not affected  by
systematics  effects  like,   e.g.   instrumental  or  the  background
polarisation  of  the  supernova  remnant.   The  \fors\  instrumental
polarisation has  been carefully characterized by  Patat \& Romaniello
(2006) who found a significant  spurious polarisation (at the level of
$\approx 1.5  \%$) close to the edges  of the CCD, while  close to the
field center is of the order  0.03\,\%, or less.  Since our target has
been positioned at the center of  the CCD, our measurement is not affected
by  the  instrumental polarisation.   We  have  finally evaluated  the
background polarisation produced by the supernova remnant by repeating
our measurements on  several objects of different brightness  close to the
pulsar position,  and we have found  that its effect  is probably less
than  1\,\%, i.e.   well within  our  global error  budget.  A  better
characterization  of  the background  polarization  introduced by  the
supernova remnants  will be possible only  when higher signal-to-noise
data  are  available.   While  our  measured  polarisation  degree  is
consistent with the one of Wagner \& Seifert (2000), who reported $\pl
=  8.5$\,\%, our associated  error turns  out to  be much  larger with
respect to their quoted 0.8\,\%.   The source of the large discrepancy
in the error estimate is  not clear, although we speculate that Wagner
\&  Seifert  underestimated  the  error  due to  the  neglect  of  the
background subtraction contribution to  the global error budget, since
0.8\,\% is  approximately the error  that we would  obtain considering
the photon  noise as  the only source  of measurement  indetermination. \\
From  the  values  of  $\pu'$  and  $\pq'$ we  have  computed  the  linear
polarization position  angle $\theta$, defined  as the angle  from the
North Celestial meridian to the major axis of the polarisation ellipse
(Landi  Degl'Innocenti et  al.  2007).   For $\pq'  > 0$,  the correct
expression to compute $\theta$ is (Landi Degl'Innocenti et al. 2007):

$$ \theta  = {1  \over 2} \arctan  \left({\pu' \over \pq'}  \right)  \;\;
\eqno{(4)} $$

\noindent
where $\theta$ counted eastward from the North Celestial meridian. For
 our  values of  $\pu'$ and  $\pq'$ we  obtain $\theta  =  -34^\circ \pm
 11^\circ$ or,  equivalently, $\theta = 146^\circ  \pm 11^\circ$.  The
 error  on $\theta$  has been  computed  using Eq.~(7)  of Fossati  et
 al. (2007).  We  have checked for a possible  coincidence between the
 pulsar  phase-averaged  optical  linear polarisation  position  angle
 $\theta$ and the position angle $\theta_X$ of the axis of symmetry of
 the X-ray arcs and jets  detected around the pulsar by \chan\ (Pavlov
 et  al.  2001).  These  structures are  most commonly  interpreted in
 terms of  particle outflows  from the pulsar  equatorial wind  and of
 collimated  outflows along  the  pulsar {\em  spin}  axis (Pavlov  et
 al. 2001;  Helfand et al. 2001). Alternatively,  they are interpreted
 in terms  of particle outflows along the  pulsar's magnetosphere open
 field  lines  and  of  collimated  outflows  along  the  pulsar  {\em
 magnetic}  axis  (Radhakrishnan   \&  Deshpande  2001;  Deshpande  \&
 Radhakrishnan 2006).  For  $\theta_X$ different estimates exist, e.g.
 $310^\circ  \pm 1.5^\circ$  (Helfand  et al.   2001), $307^\circ  \pm
 2^\circ$ (Pavlov et al.  2001), and $310.63^\circ \pm 0.07^\circ$ (Ng
 \& Romani 2004).  It has also been noted (e.g.  Helfand et al.  2001)
 that  $\theta_X$   apparently  coincides  with   the  position  angle
 $\theta_{\mu}$  of  the pulsar  proper  motion.  For  $\theta_{\mu}$,
 based on  \hst\ observations Caraveo  et al. (2001b)  give $307^\circ
 \pm  2^\circ$, while  Dodson  et  al.  (2003)  with  the \vlbi\  give
 $301^\circ \pm 1.8^\circ$.  Interestingly, our measurement of $\theta$ is
 not  very different (within  $\approx 2  \sigma$) from  the available
 measurements of  $\theta_X$ and $\theta_{\mu}$,  suggesting an alignment
 between  the  optical  linear  polarisation direction,  the  axis  of
 symmetry of  the X-ray arcs and  jets, and the  pulsar proper motion.
 Although the  chance occurrence probabilities of  this alignment are
 not negligible, it  is tantalising to speculate about  it as a tracer
 of the  connection between  the pulsar's magnetospheric  activity and
 its dynamical interactions with the surrounding medium.  More precise
 measurements of  the pulsar  optical polarisation, possibly  supported by
 still to come polarisation measurements in X-rays, will hopefully provide
 a   further  observational   handle  for   more   robust  theoretical
 speculations.

\section{Discussion}

We have  then compared our  measurement of the Vela  pulsar phase-averaged
optical polarisation  degree with  the expectations of  various pulsar
magnetosphere models.  We have used  the polarisation code of  Dyks et
al.~(2004,   hereafter   DHR04)  which   can   calculate  the   linear
polarisation degree  for the  emission region of  the outer  gap model
(Cheng  et al.~1986;  Romani  \& Yadigaroglu  1995)  and its  two-pole
version  (Dyks \&  Rudak 2003,  hereafter DR03).   The  code carefully
takes into  account the macroscopic properties of  the emission region
(3D  spatial  extent,  differential  aberration,  propagation  delays,
cumulation of  radiation emitted from different parts  of the emission
region  at  the  same  pulse  phase)  but it  ignores  much  of  micro
physics.  For instance,  the  intrinsic polarisation  degree \pli\  of
emitted radiation is assumed to be the same within the entire emission
region (it  is an  input parameter of  the code) and  the polarisation
angle is given  by projection of ``bulk" electron  acceleration on the
observer's sky.  The acceleration is only determined  by the curvature
of magnetic field lines and the corotation of pulsar magnetosphere (no
gyration component).  Within the emission region of a given model, the
radiation power  is independent  on the altitude  (the same  number of
photons  is emitted  per centimetre  of electron's  trajectory  in the
observer's  frame).   In the  magnetic  colatitude  the emissivity  is
limited  to  a narrow  range  of  the  magnetic field  line  footprint
parameter.

\subsection{Comparison with the outer gap model}

For the  outer gap model we  assumed that the  emission region extends
from  the null  charge  surface  (Romani \&  Yadigaroglu  1995) up  to
$r_{\rm max}=1.5~ R_{\rm lc}$ with $\rho_{\rm max} = 0.95 ~R_{\rm lc}$
($R_{\rm lc}  = c P_{\rm  rot} /2 \pi$  is the light  cylinder radius,
where $P_{\rm  rot}$ is  the pulsar rotational  period and $c$  is the
speed  of light), we  considered magnetic  field lines  with footprint
parameter  $0.75  \la r_{\rm  ovc}  \la  0.95$,  and we  weighted  the
emission  in magnetic colatitude  by a  Gaussian centred  at $r^0_{\rm
ovc}=0.85$ ($\sigma=0.05$).   All the previous  parameters are defined
in DHR04.   For the Vela pulsar emission  geometry different estimates
exist,  e.g.   $\alpha=71^\circ$,  $\zeta=65^\circ$ (Radhakrishnan  \&
Deshpande 2001), where $\alpha$  is the dipole inclination and $\zeta$
is the  viewing angle (see,  e.g. Fig.~1 of  DR03), $\zeta=63.6^\circ$
(Ng \& Romani 2004) from  two different geometrical models of the Vela
pulsar  X-ray  nebula;  $\alpha\simeq70^\circ$,  $\zeta\simeq61^\circ$
(DR03),  from   the  two-pole  model  of   the  $\gamma$-ray  profile;
$\alpha\sim55^\circ$,  $\zeta \sim  49^\circ$  (Johnston et  al.~2001)
from  a fit  to  the position  angle  curve at  $0.6$  and $1.4$  GHz;
$\alpha\sim137^\circ$, $\zeta \sim 143.5^\circ$ (Johnston et al.~2005)
from  a fit at  $1.4$ GHz;  $\zeta=53.8^\circ$ (Krishnamohan  \& Downs
1983) from a  fit of  the position  angle curve at  $2.3$ GHz  and for
$\alpha=60^\circ$.   In  the  following,  we  will  assume  $\alpha  =
71^\circ$ as a representative value.

\begin{figure}
   \centering
    \includegraphics[width=0.5\textwidth]{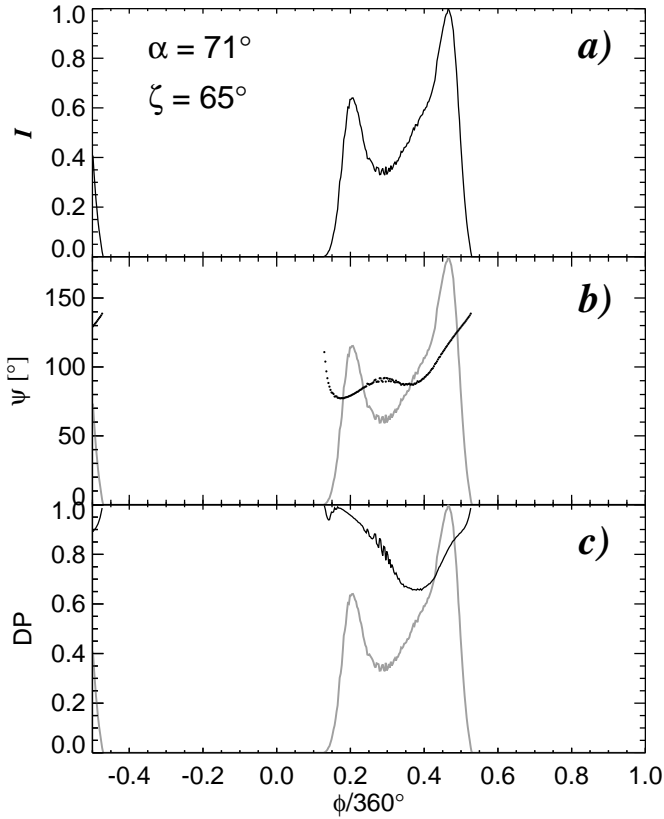}
      \caption{Phase-resolved  polarisation  properties  of  the  Vela
pulsar calculated for a  simplified outer gap calculation (DHR04) with
the  emission region  parameters described  in the  text  and assuming
$\alpha=71^\circ$ and $\zeta=65^\circ$. The  panels show: {\bf a)} the
pulse  profile;  {\bf b)}  the  modelled  polarisation position  angle
$\psi$;  {\bf  c)}  the   depolarisation  factor  (see  text  for  the
definitions).  The radio pulse is expected near phase zero.  One and a
half rotation  period is  shown.  The light  curve is over  plotted in
panels b and c for reference (grey line).  }
         \label{phresolved}
   \end{figure}

Fig.~\ref{phresolved}  shows the  modelled Vela  pulsar  optical pulse
profile  and  the   phase-resolved  polarisation  properties  for  the
representative case of  $\alpha=71^\circ$ and $\zeta=65^\circ$ and for
the same emission region parameters assumed above.  As it is seen from
the top  panel, this set  of parameters can reasonably  well reproduce
the  observed optical  pulse profile  of  the Vela  pulsar (see,  e.g.
Gouiffes 1998), with two peaks separated by $\sim0.25$ in phase. Thus,
we are  confident that they provide  a good approximation  of the Vela
pulsar's emission  geometry.  The two  lower panels show  the modelled
polarisation position angle $\psi$, measured from the projected pulsar
rotational  axis,  and   the  depolarisation  factor  ${\rm  DP}\equiv
\plm/\pli$,  defined as  the ratio  between the  modelled polarisation
degree   $\plm$  at   the  observer's   location  and   the  intrinsic
polarisation degree  of the emitted radiation $\pli$.   From our model
we have derived the phase-averaged value of the depolarisation factor,
where the  average is  done over the  Stokes parameters.   We obtained
${\rm DP} =0.68$,  a typical value for many  sets of model parameters.
From  this value  we have  then computed  the  phase-averaged modelled
polarisation  degree  $\plm$ for  different  values  of the  intrinsic
polarization $\pli$.

\begin{figure}[t]
   \centering
   \includegraphics[width=0.5\textwidth]{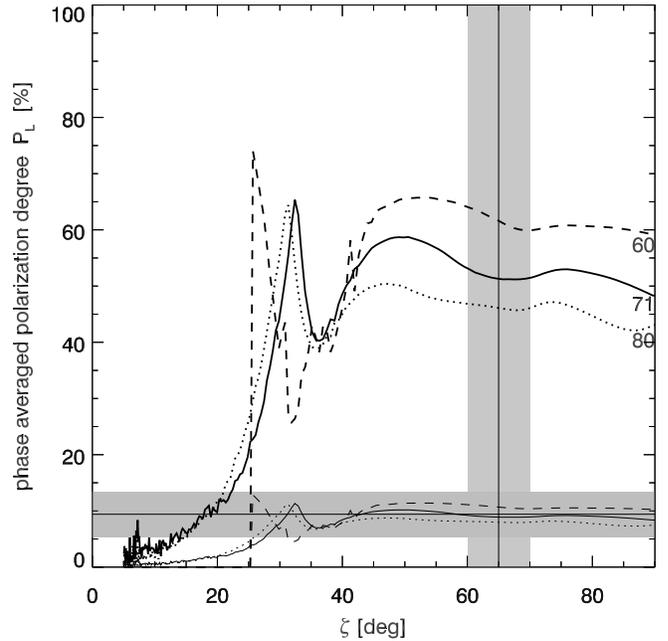}
      \caption{Modelled  phase-averaged   linear  polarisation  degree
\plm\ of  the Vela pulsar as  a function of the  viewing angle $\zeta$
for  a simplified  outer  gap  calculation (DHR04)  and  for a  dipole
inclination  $\alpha=71^\circ$ (solid).   As a  reference,  curves for
$\alpha=60^\circ$  (dashed) and  $\alpha=80^\circ$  (dotted) are  also
shown.   The thick  and  the  thin lines  correspond  to an  intrinsic
optical polarisation  $\pli=75\%$ and $\pli=13\%$,  respectively.  The
range  of viewing  angles $\zeta$  derived with  various  methods (see
text)  is  marked  with   the  vertical  shaded  band.   The  measured
phase-averaged polarisation  degree $\pl$ (this work) is  shown as the
horizontal grey band.  }
         \label{phaveraged}
   \end{figure}

\begin{figure}
   \centering
   \includegraphics[width=0.5\textwidth]{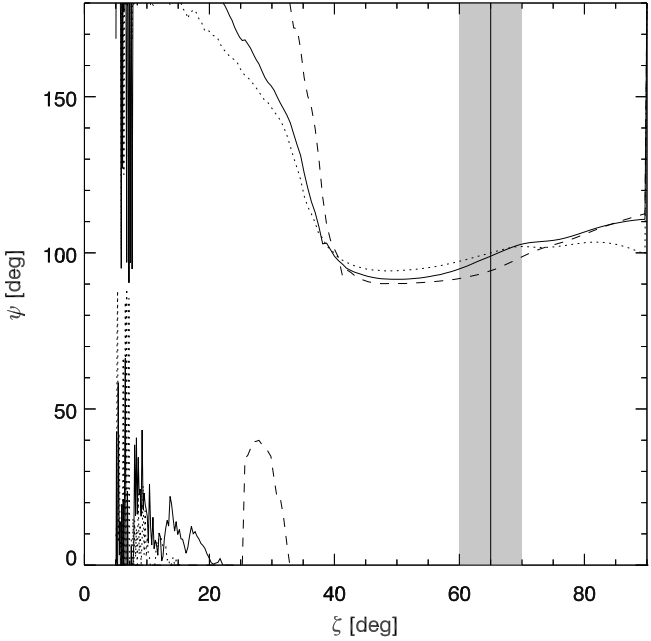}
      \caption{Phase-averaged  modelled  position   angle  $\psi$  as  a
function of  the viewing  angle $\zeta$ for  the same range  of dipole
inclinations $\alpha$ as in Fig.  ~\ref{phaveraged}.  The range of viewing angles $\zeta$
derived with  various methods (see  text) is marked with  the vertical
shaded band.  }
         \label{psi}
   \end{figure}

\noindent
Fig.    ~\ref{phaveraged}   presents   the   modelled   phase-averaged
 polarisation degree  $\plm$ of the Vela  pulsar as a  function of the
 viewing  angle $\zeta$  for  two different  values  of the  intrinsic
 optical polarisation $\pli$ (see caption).  As it is seen, for $\zeta
 \simeq 65^\circ$ the modelled polarisation degree is much larger than
 the  observed one  if  $\pli=75\%$, while  consistency would  require
 $\pli$ to be as low as $13\%$.  Fig.~\ref{phaveraged} also shows that
 for large  $\alpha$ the modelled  polarisation degree \plm\  does not
 vary  significantly for  viewing angles  $\zeta \ga  40\%$.   Thus we
 find, in agreement with Chen  et al.  (1996), that the phase-averaged
 polarisation \pl\ for the outer  gap models is usually more sensitive
 to  the intrinsic  properties of  the emitted  radiation than  to the
 global geometric parameters ($\alpha$, $\zeta$).  While this makes it
 difficult to  constrain $\alpha$ and  $\zeta$ from the  comparison of
 the modelled and observed \pl\,  it also implies that our conclusions
 are less affected by the uncertain knowledge of the pulsar's emission
 geometry. Thus, our simulation based  on the standard outer gap model
 overpredicts the value of the phase-averaged polarisation, unless the
 intrinsic  polarisation is much  lower than  expected.  This  is most
 likely due  to the fact  that the code  of DHR04 does not  takes into
 account the convolution of the single electron emission beam with the
 structure  of the  emission region/magnetic  field, which  can  be an
 important source of depolarisation,  cf.~ Fig.~4b and Fig.~6c in Chen
 et al.  (1996) and eq.~(42)  in Epstein (1973).\\ For large $\alpha$,
 and for $\zeta  > 30^\circ$ the two-pole caustic  model (DR03), which
 includes  the strong  emission from  below the  null  charge surface,
 predicts roughly two  times smaller \pl\ than the  standard outer gap
 model.  However,  the two-pole models (including  the ``extended slot
 gap"  of Muslimov  \& Harding  2004) can  not reproduce  the observed
 optical  pulse shape  of the  Vela  pulsar.  This  may just  indicate
 different emission  regions for the optical and  for the $\gamma$-ray
 band, where the two-pole geometry performs quite well.

We have then derived the phase-averaged modelled polarisation position
angle $\psi$,  where, as before, the  average is done  over the Stokes
parameters.   Fig.  ~\ref{psi}  presents  the modelled  phase-averaged
polarisation position angle $\psi$ as  a function of the viewing angle
$\zeta$ for different values of the dipole inclination $\alpha$.  Only
one  set of  curves  is drawn  since  $\psi$ does  not  depend on  the
intrinsic polarization degree  $\pli$.  As it is seen,  for $\zeta \ge
40^\circ$  the value  of $\psi$  depends very  little on  $\alpha$ and
$\zeta$.  We  note that  the actual value  of modelled  position angle
$\psi$ depends on which emission mechanism is actually responsible for
the  optical  emission   (e.g.   synchrotron  radiation  or  curvature
radiation).   In the  case of  Fig.  ~\ref{psi},  $\psi$  was computed
assuming  that  the   emitted  waves,  after  appropriate  microscopic
averaging  that  is  not  modelled,  are  polarized  parallel  to  the
projected   bulk  acceleration  of   electrons.   This   direction  of
polarization is  characteristic of curvature  radiation (Jackson 1962)
and it is  parallel to projected trajectory of  electrons' bulk motion
(with  no   gyration).   Instead,   if  the  radiation   is  polarized
perpendicular to the direction assumed in our code, as is the case for
the synchrotron radiation (Rybicki  \& Lightman 1979) one must correct
$\psi$  by  subtracting  $90^\circ$.  So  the value  of  the  modelled
position  angle   should  be  understood   "modulo  90$^\circ$".   For
instance, for $\zeta  \sim 65^\circ$ the modelled values  of $\psi$ is
either  $\sim  98^\circ$ or  $8^\circ$  for  polarization parallel  or
perpendicular to the  projected bulk acceleration, respectively.  This
means that the modelled  phase-averaged polarisation vector is roughly
perpendicular or aligned to the  pulsar rotation axis.  For $\zeta$ on
the other hemisphere ($\zeta' = 180^\circ-\zeta$) the derived position
angle   $\psi$   would    just   change   sign   i.e.    $\psi(\zeta)=
-\psi(\zeta')$, see Table 1 in DHR04.

A  striking result  visible in  Fig.  ~\ref{psi}  is the  proximity of
$\psi$  to  $\sim90^\circ$ within  the  large  range  of $\alpha$  and
$\zeta$.   To  understand  it   we  have  recalculated  the  model  of
Fig.~\ref{phresolved}  with   various  kinematic  effects  temporarily
removed from our code.  We  have run the code successively without the
sweepback  of magnetic field  lines, without  centrifugal acceleration
and without aberration.   This exercise has shown that  it is the last
effect (the  aberration of  photon emission direction  associated with
transition from the corotating frame to the observer's frame) which is
the  most  important  in  fixing  $\psi$  near  $\sim90^\circ$.   This
contribution of corotational velocity to the emission direction in the
observer's frame  is large because  the optical emission of  the outer
gap model  originates from altitudes that are  a considerable fraction
of the light cylinder radius.

  It is now interesting  to compare the modelled polarisation position
angle  $\psi$ with  the observed  one  $\theta$. However,  we have  to
caution  that a  straight comparison  between  the two  angles is  not
possible  since  the former  is  measured  from  the projected  pulsar
rotational  axis,  while  the  latter  is measured  from  North.   The
modelled and observed polarisation position angles are thus related by
the equation $\psi = \theta  - \psi_0$, where $\psi_0$ is the position
angle of the  pulsar's projected rotational axis defined  in the usual
way,  i.e.  measured eastward  from North.   Thus, equating  $\theta =
146^\circ \pm 11^\circ$ (see Sect.~2.3) and $\psi$ allows for a direct
measurement of $\psi_0$.  For  $\psi = 98^\circ \pm 5^\circ$ (polarization
parallel to  the bulk acceleration)  we obtain $\psi_0 =  48^\circ \pm
12^\circ$.  Instead,  for $\psi  = 8^\circ \pm  5^\circ$ (polarization
perpendicular to bulk acceleration)  we obtain $\psi_0 = 138^\circ \pm
12^\circ$.  The first case  implies that the pulsar projected rotation
axis is almost perpendicular to the axis of symmetry of the X-ray jets
and arcs, while  the second case implies that it  is almost aligned to
them (see Sect.~2.3).  If we assume that this geometry is the favoured
one (e.g.  Helfand et al.  2001;  Pavlov et al.  2001), then $\psi_0 =
138^\circ \pm 12^\circ$, $\psi =  8^\circ \pm 5^\circ$ and the optical
radiation actually is polarised perpendicular to the bulk acceleration
assumed in our  code. This is consistent with  a synchrotron radiation
origin of  the optical emission.  The  other way around,  if we assume
that  the  optical  emission  is  due to  synchrotron  radiation,  our
observed  polarisation  position angle  would  imply  that the  pulsar
rotation axis is aligned with the  axis of symmetry of the X-rays jets
and arcs.

Thus, our  results would imply  that there is a  substantial alignment
between the  optical polarisation position angle,  the pulsar rotation
axis and proper  motion vector, and the axis of  symmetry of the X-ray
nebula.  Evidence  for a  similar alignment has  been also  found from
phase-resolved radio  polarisation measurements  of a number  of radio
pulsars, including Vela (Johnston et al. 2005).

\subsection{Comparison with other models}

 We  have  qualitatively  evaluated  other  models  by  comparing  the
predicted vs.  observed polarisation properties published for the Crab
pulsar  which has a  very similar  geometry to  the Vela  pulsar.  The
viewing angle  $\zeta \simeq 60^\circ$ for  both Crab and  Vela (Ng \&
Romani 2004)  and the  dipole inclination $\alpha$  is believed  to be
large in both cases (e.g.~Rankin 1990).  Furthermore, as we have shown
above, the  polarisation properties are found to  depend rather weakly
on $\alpha$ and  $\zeta$.  \\ One of the  more sophisticated models is
the striped wind model (Petri  \& Kirk 2005). However, this model does
not    account   for    magnetic   field    irregularities    in   the
emission/reconnection   region   and   this   is  probably   why   the
phase-averaged polarisation  degree calculated for the  Crab pulsar is
much  larger   (19-31\%,  Petri  \&  Kirk  2005)   than  the  observed
one\footnote{In   the  case   of   the  Crab   pulsar,  the   observed
phase-averaged  polarisation degree  is equal  to $6.2\%$.   After the
``DC" component is subtracted  (Kellner 2002; Kanbach et al.~2005) the
phase averaged  \pl\ becomes  $3.7\%$.}.  A version  of the  outer gap
model by Chen et al.  (1996)  is also notably advanced, as it includes
convolution  of  the  single   electron  emission  beam  with  spatial
distribution,   electron   energy   distribution   and   pitch   angle
distribution. The neglected magnetic field irregularities are probably
less important  in this  case because the  emission originates  from a
region  well inside  $R_{\rm  lc}$.  However,  the assumed  simplified
geometry of the emission region (2D, no radial extent) seems not to be
adequate as it does not  allow to reproduce the observed pulse profile
of the Crab  pulsar. Accordingly, this model does  not account for the
depolarisation effect produced by  the cumulation of radiation emitted
from different altitudes.  Using Fig.~6  in Chen et al.~(1996) one can
calculate a \emph{partial}  phase-averaged polarisation degree that is
limited to the phase range  within $\pm44^\circ$ around the maximum of
main pulse.  The  result ($9.4$\%) is two times  larger than the value
observed for the  Crab pulsar within the same  phase range ($5.1$\% or
$4.2$\% if the DC emission is excluded).

\section{Conclusions}

We have  studied the  Vela pulsar phase-averaged  optical polarisation
properties  by reanalysing  the data  set  used by  Wagner \&  Seifert
(2000). The  fraction of  linear, phase-averaged polarisation  ($\pl =
9.4\% \pm 4\% $) is qualitatively  similar to what they found but with
a larger error  that we attribute to a  more realistic error analysis,
thus   underscoring   the    need   for   observations   with   higher
signal-to-noise ratio.   In addition, we  have measured for  the first
time  the  optical  linear  polarisation  position  angle  ($\theta  =
146^\circ \pm  11^\circ$) finding that  it is apparently  aligned with
the axis of symmetry of the  X-ray arcs and jets and with the pulsar's
proper  motion  vector.   We   have  compared  our  results  with  the
expectations  of  the   outer  gap  as  well  as   of  various  pulsar
magnetosphere models.  Interestingly, for  the outer gap model we have
found  that  for  synchrotron  optical emission  the  observed  linear
polarisation  position angle  $\theta$  implies that  also the  pulsar
rotational axis is aligned with the axis of symmetry of the X-ray arcs
and jets.   For all models  we found that the  computed phase-averaged
polarisation degree  is typically much  larger than the  observed one.
The  most  probable reason  for  this is  that  these  models are  too
idealised  and  do  not   account  for  some  depolarisation  effects.
Measurements of  phase-averaged polarisation of pulsars  can thus help
to  identify  the  weaknesses  of  various models  and  stimulate  the
development of  the complex numerical codes required  to calculate the
expected   polarisation   properties.    More   optical   polarisation
observations of pulsars,  now in progress with the  \hst, will help to
provide  the   very  much  needed  observational   grounds  to  settle
theoretical models.

\begin{acknowledgements}
RPM  warmly thanks  the  ESO/Chile Scientific  Visitors Programme  for
supporting his  visit at the ESO  Santiago Offices where  most of this
work was finalised.   JD and AS acknowledges the  support from the KBN
grant  2P03D.004.24.  AS  also acknowledges  support from  a Deutscher
Akademischer  Austausch  Dienst   (DAAD)  fellowship.   We  thank  the
anonymous  referee for  his/her useful  comments which  contributed to
improve the quality of the paper.
\end{acknowledgements}

\end{document}